\documentstyle[12pt,a41]{article}

\begin{document}

\begin{center}
{\LARGE\bf Fast simulation of a HERA--like detector}

\vspace{1cm}
{J.\ G.\ Contreras$^a$}

\vspace*{1cm}
{\it $^a$Universit\"at Dortmund, Institut f\"ur Physik, D--44221
Dortmund, Germany}\\

\vspace*{2cm}

\end{center}

\begin{abstract}
  A set of FORTRAN routines to perform a fast simulation of a
  HERA--like detector including the smearing of the $z$ coordinate of
  the interaction vertex, the simulation of a tracker system, of an
  electromagnetic and  of  a hadronic calorimeter is presented.
\end{abstract}

\section{Introduction}

\vspace{1mm}
\noindent
One of the main goals of this workshop has been to study the
measurability, under realistic experimental conditions, of the
different observables from spin asymmetries proposed in
\cite{fut_at_hera}. To this end a fast simulation of a HERA--like
detector has been implemented and will be described here. It consist
of one function and four subroutines, written in standard FORTRAN,
which can be called directly by the user: \texttt{z\_vtx, ele\_det, had\_det,
track\_det}.  It also includes two private functions: \texttt{thvtx,
gauss\_vec}.  The sources can be found in \cite{www}.

The package covers the simulation of the smearing in the
$z$--coordinate of the interaction vertex\footnote{The positive $z$
axis is defined by the direction of the incoming proton beam}, the
simulation of an electromagnetic and a hadronic calorimeter, and that
of a tracker system.

The effects taken into account were the following. The acceptance of
the detectors in the polar direction\footnote{angle with respect to
the positive $z$ axis} (It was assumed that the coverage in the
azimuthal direction amounts to 2$\pi$). An effective distribution of
dead material between the interaction vertex and the trackers and/or
the calorimeters. Smearing of the energy (the transverse momentum for
the tracks) and angles of the generated particles. The possibility to
alter the absolute energy scale to perform systematic studies.

The efficiency to trigger and select a given event have not been
simulated and must be taken from somewhere else. All parameters used
in these routines have realistic values, and are quite similar to the
equivalent values used by the H1 \cite{h1_det} and ZEUS
\cite{zeus_det} collaborations at HERA.

\section{The interaction vertex and the auxiliary routines}

\vspace{1mm}
\noindent
At HERA the bunches have a big size in the $z$ direction compared to
their transverse size. Thus the $z$ coordinate of the interaction
vertex is distributed over several centimetres around the nominal
interaction point \cite{hera}. The function
\begin{quote}
\texttt{real function z\_vtx(z\_0, z\_sigma)\\
   real z\_0, z\_sigma}
\end{quote}
returns a random value for the $z$ coordinate of the interaction
vertex assuming a Gaussian distribution of mean value \texttt{z\_0}
and variance \texttt{z\_sigma}, where the units are in centimetres. If
\texttt{z\_sigma = 0} then \texttt{z\_sigma=10.0} is assumed. This
function has to be called first and only once per event.

To obtain the needed random numbers used by the simulations routines a
generator of a vector of Gaussian numbers have been implemented
\begin{quote}
\texttt{subroutine gauss\_vec(g,n)\\
 real g(*) \\
 integer n}
\end{quote}
It takes as input an even integer value \texttt{n} and returns the
vector of \texttt{n} Gaussian numbers in \texttt{g}.

The dead material correction factors are stored as a function of the
polar angle $\theta$ assuming an interaction vertex at $z=0$.
Therefore, given the smearing of the vertex, it is necessary to
perform a transformation to to find the appropriate correction factor.
This occurs internally via the function \texttt{thvtx} which is
automatically called when needed and the user does not have to worry
about the details.

\section{The tracker system}

\vspace{1mm}
\noindent
The simulated tracker system consist of a central part covering the
polar range $20^\circ<\theta<160^\circ$ and a forward part with
acceptance $7^\circ<\theta<25^\circ$. The trackers simulate only
particles with a transverse momentum bigger than 200 MeV and have a
reconstruction efficiency of 95\% and 70\% percent respectively.
Each section has its own sets of parameters to smear the transverse
momentum of the particles as well as their polar angle. All the values
used for the different parameters are in accordance with those quoted
in \cite{h1_track, zeus_track}. 

This routine has to be call inside of a loop of all the charged
hadrons generated in the event and has following syntax
\begin{quote}
\texttt{subroutine track\_det(z\_vtx,vin,vout)\\
      real z\_vtx, vin(4), vout(4)}
\end{quote}
where the input is the $z$ coordinate of the interaction vertex for
the current event \texttt{z\_vtx}, and the four momentum
($p_x,p_y,p_z,E$) of the charged particle \texttt{vin}. The output is
the simulated four momentum \texttt{vout}.
  
\section{The calorimeter}

\vspace{1mm}
\noindent
The calorimeter consist of an electromagnetic part to simulate the
measurement of the scattered electron in DIS and a hadronic part to
simulate the measurement of the hadronic final state.

Both calorimeters consist of a central and a backward part. They cover
the polar range from 4$^\circ$ to 177$^\circ$, whereas the central and
backward part meet at 152$^\circ$. An optional acceptance in the range
$10^\circ<\theta<170^\circ$ can be used to take into account an
eventual upgrade of the HERA machine \cite{upgrade}. The complete
azimuthal angle $\phi$ is covered for both calorimeters. 

The influence of the dead material before the calorimeters and the
losses due to geometric acceptance at the edges and in the region
between the central and backward parts was modelled with correction
factors which depend on the polar angle of the particle to be
simulated. For the hadronic calorimeter there are 80 of these
factors. For the electromagnetic calorimeter are two sets of 24
factors. Which of the two sets is used depends on the energy of the
scattered electron.

Smearing of the energy and angles of the particles measured with the
hadronic calorimeter was applied using the following realistic factors
\cite{lar}. The energy resolution of the
backward part is $\sigma/E=0.56$, and of the central part
$\sigma/E=0.46/\sqrt{E}\oplus0.73/E\oplus0.026$, where the
energy is given in GeV. For the angular resolutions in mrad
$\sigma_\theta = 50$ and $\sigma_\phi=90$ were used.

For the measurement of the scattered electron following values have
been used (see for example \cite{spacal}). In the backward part 
$\sigma_{cent}/E=0.071/\sqrt{E}\oplus0.01$,
$\sigma_\theta = 2$ and $\sigma_\phi=6$ were applied. In the central
part $\sigma_{cent}/E=0.13/\sqrt{E}\oplus0.05$,
$\sigma_\theta = 9$ and $\sigma_\phi=30$ were used.
Again the energies are given in GeV and the angles in mrad.

The absolute energy scale of the calorimeters, which
is one of the main sources of uncertainty for the different
measurements, was assumed to be known within 4\% and 10\% for the
central, respectively for the backward, part of the hadronic
calorimeter. For the electromagnetic section the values were 3\% and
1\%.

The routine for the electromagnetic section 
\begin{quote}
\texttt{subroutine ele\_det(z\_vtx, vin, vout, en\_sc ) \\
      real z\_vtx, vin(4), vout(4) \\
      integer en\_sc }
\end{quote}
has to be called once per event. The input is the $z$ coordinate of the
interaction vertex \texttt{z\_vtx}, the four momentum of the generated
scattered electron \texttt{vin}, and a flag \texttt{en\_es} to govern
the absolute energy scale of the calorimeter: if it is positive
(negative) the energy will be up(down)scaled, otherwise it will remain
unaltered. The four momentum of the simulated scattered electron is
returned in \texttt{vout}. The hadronic section 
\begin{quote}
\texttt{subroutine had\_det(z\_vtx,vin, vout, en\_sc) \\
real z\_vtx, vin(4), vout(4)\\
integer en\_sc}
\end{quote}
 has to be called inside a loop of all the particles of the
event except the scattered electron. The variables act as in the
routines already presented.

\section{Conclusions}

\vspace{1mm}
\noindent A set of FORTRAN routines to simulate in a time effective
way the behaviour of a HERA--like detector to study the measurability
of spin asymmetries at HERA has been implemented. They are described
here. It includes the smearing of $z$ coordinate of the interation
vertex, the simulation of a tracker system as well as that of an
electromagnetic and a hadronic calorimeter. Effects included are
acceptance of the detector, correction for dead material and smearing
of the four momentum of the simulated particle.

\end{document}